# GAMMA-RAY BURSTS AND COSMOLOGY

Jay P. Norris

Laboratory for High Energy Astrophysics,
NASA/Goddard Space Flight Center, Greenbelt, MD, USA 20771

**Abstract**. The unrivalled, extreme luminosities of gamma-ray bursts (GRBs) make them the favored beacons for sampling the high redshift Universe. To employ GRBs to study the cosmic terrain – e.g., star and galaxy formation history – GRB luminosities must be calibrated, and the luminosity function versus redshift must be measured or inferred. Several nascent relationships between gamma-ray temporal or spectral indicators and luminosity or total energy have been reported. These measures promise to further our understanding of GRBs once the connections between the luminosity indicators and GRB jets and emission mechanisms are better elucidated. The current distribution of 33 redshifts determined from host galaxies and afterglows peaks near $z \sim 1$, whereas for the full BATSE sample of long bursts, the lag-luminosity relation predicts a broad peak $z \sim 1$–4 with a tail to $z \sim 20$, in rough agreement with theoretical models based on star formation considerations. For some GRB subclasses and apparently related phenomena – short bursts, long-lag bursts, and X-ray flashes – the present information on their redshift distributions is sparse or entirely lacking, and progress is expected in Swift era when prompt alerts become numerous.

## 1. COSMIC TERRAINS

It is now almost seven years since the discovery of GRB afterglows, which led immediately to the understanding that the sources of long (duration > 2 s) GRBs lie at cosmological distances. When considering the uses of GRB for cosmological inquiry, it is instructive to keep in mind lessons from the AGN experience, in which jets, relativistic beaming, and viewing angle determine to a large degree the observed luminosity, as in GRBs. Putting GRBs into context, after decades of work the AGN luminosity distribution is still a matter of complex study at higher redshifts, $z > 2$, and must be inferred by indirect arguments (Schirber & Bullock 2003). Partly, the uncertainties are attributable to nondetection of the lower luminosity objects at higher redshifts, but also systematic effects often make for difficult calibration. As an example, one of few effects relevant to both AGN and GRB luminosity measurements is weak gravitational lensing.

While the detailed understanding of the GRB luminosity distribution is still developing, GRBs are widely seen as potentially excellent markers of star formation at early epochs. At redshifts beyond $z \sim 2$, star formation rates (SFR) are highly uncertain and major determinants for initial stellar masses – such as metallicity and the particulars of cloud fragmentation/coalescence processes – are topics of many recent theoretical papers. Mackey, Bromm, & Hernquist (2003) discuss SFR governed by molecular and atomic hydrogen cooling at $z \sim$ 10–35 (spanning the bounds on the start of the reionization epoch), giving way to multi-phase star formation at later times. During the earlier era, the very massive population III stars formed, driving high SN rates per unit volume, presumably many generated with conditions required to make GRBs.

The fact that GRBs tend to originate in galaxies with high SFR was demonstrated by Berger et al. (2003a) from a study of radio emission from GRB hosts. The radio channel reveals that the hosts have approximately an order of magnitude higher SFR than optical results suggest, and have significantly bluer colors than pre-selected sub-mm radio galaxies inhabiting a redshift range comparable to the current median redshift for GRBs, $z \sim 1$. Detectability of GRBs and their afterglows to the highest redshifts where we may expect the earliest star formation was originally discussed by Lamb & Reichart (2000).

The ingredient essential to relating GRBs to SFR is dependence of type of stellar death on the initial stellar configuration – initial mass, metallicity, and rotation rate. Heger et al. (2003) discuss how massive stars expire, including how those with more than $\sim 40$ solar masses and sufficient metallicity may evolve into collapsars and produce GRBs. Fryer & Meszaros (2003) discuss aspects of stellar collapse dependent on remnant mass and rotation rate in the context of neutrino-driven explosions, and predict the interval between collapse and ensuing GRB explosion in terms of black hole spin rate (10s of seconds to $\sim 1$ month).

Another possible cosmological application of GRBs discussed recently is measurement of the fundamental parameters of the Universe. While the cosmic microwave background (Bennett et al. 2003) and Type Ia supernovae (SNe) have been used to realize mutually reinforcing accuracies to the level of a few percent for most parameters, the evolution with cosmic time of "w" – the equation of state for the dark energy content – is not yet well constrained (Turner 2001). Recent SN search efforts have concentrated in the regime z ~ 0.5, where the turnaround from deceleration to acceleration occurs by virtue of the presumed growing dominance of dark energy. At redshifts greater than unity, the deceleration epoch is not yet well defined in Hubble plots. Possible systematics for SNe Ia measurements still require attention, including metallicity and stellar age variation with redshift, and weak lensing beyond z ~ 0.5, and sample variations of order 0.05 magnitude still appear in different treatments (e.g., Tonry et al. 2003; Blakeslee et al. 2003). In principle, GRB Hubble plots could make significant contribution to the measurement of w(t) since GRB sources inhabit the relevant redshift regimes and GRBs are easily detected.

However, in addition to the redshift luminosity, the physical luminosity must be measured or inferred accurately as well to produce a useful Hubble plot. The next section briefly discusses nascent gamma-ray luminosity indicators and some of the problems inherent in making accurate gamma-ray measurements required for cosmology parameters.

## 2. GAMMA-RAY LUMINOSITY AND ENERGY INDICATORS

Based on handfuls of bursts with spectroscopic redshifts, several relationships between gamma-ray properties and luminosity or radiated energy have been reported. These include correlation between variability measures and gamma-ray luminosity (Fenimore & Ramirez-Ruiz 2000; Reichart et al. 2001); anti-correlation between spectral lag and luminosity (Norris, Marani, & Bonnell 2000); correlation between variability and $E_{peak}$ (Lloyd-Ronning & Ramirez-Ruiz 2002), and correlation between $E_{peak}$ and total gamma-ray energy (Amati et al. 2002). It would not be surprising if it were necessary to combine such indicators in order to realize eventually a more robust and accurate measure, since at least three intrinsic or extrinsic variables probably determine the observed luminosity: the jet profiles for Lorentz factor and mass density, and observer viewing angle. As an example of what the present-day coarse relationships predict in terms of the GRB redshift distribution, Figure 1 illustrates the distribution inferred for BATSE long bursts from the lag-luminosity relationship, once an extraneous factor of (1+z) is eliminated (from eq. 7, Norris 2002). Logarithmically binned, a broad peak extends across z ~ 1–4 with a tail to z ~ 20, in rough agreement with models based on star formation considerations where the GRB luminosity function is assumed to be time-independent (Bromm & Loeb 2002). Some aspects which may present irreducible floors to improving the systematics in such gamma-ray luminosity indicators include weak gravitational lensing, extrapolated corrections for the finite instruments bandpasses, and the complexity of GRB time profiles.

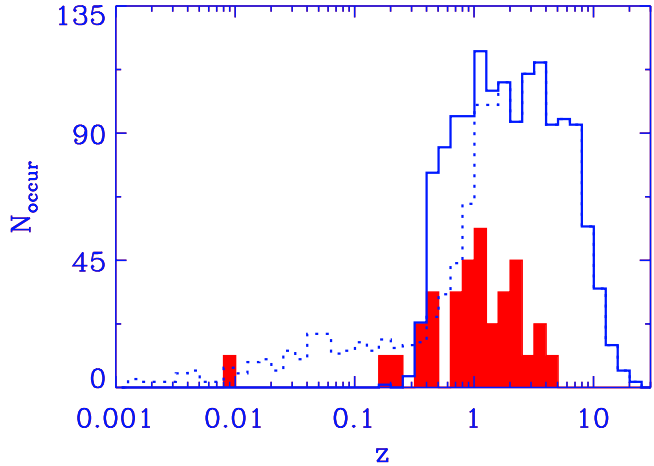

Fig. 1. Distribution of redshifts inferred for 1429 BATSE long bursts, assuming one-branch (solid) and two-branch (dotted) lag-luminosity relations (Band, Norris, & Bonnell 2004). Solid fill histogram (with different ordinate scale) is distribution of 33 bursts with spectroscopically determined redshifts.

3. SPARSELY DEFINED COSMOLOGIES OF GRB SUBCLASSES

Long-lag GRBs have relatively long spectral lags, ~ 1–10 s, as measured between low (~ 25–50 keV) and high (~ 100–300 keV) energy bands. This subclass is important for several reasons. It dominates the BATSE sample of long bursts below a peak flux of ~ 0.7 photons $cm^{-2}$ $s^{-1}$. Many such bursts manifest most of their emission in one pulse, being good examples of the canonical fast rise exponential decay. Thus they are often sufficiently simple to analyze that they may yield insight into the more complex, short-lag GRBs. Several groups predicted subclasses similar to the observed long-lag bursts – with ultra-low luminosities compared to most GRBs, soft spectra, and possibly long spectral lags, where these properties may be attributed to a combination of low Lorentz factor, large jet opening angle, and/or large viewing angle (Woosley & MacFayden 1999; Ioka & Nakamura 2001; and Salmonson 2001). Presently only one member of the subclass has a spectroscopically determined redshift, GRB 980425 / SN 1998bw, with $z = 0.0085$. If this burst is representative, then many long-lag bursts would lie within a distance of a few times 100 Mpc, and would represent the low end of the GRB luminosity distribution (Norris 2002). How nearby is an important question, since GRB 980425 may somehow be anomalous. Berger et al. (2003b) searched for radio emission from 33 Type Ib/c SNe, finding none as energetic as GRB 980425. The interpretation is that only hydrodynamic, non-relativistic flow is indicated for these SNe, unlike conditions expected for GRBs. Thus the general relationship between SNe and long-lag GRBs is in doubt, along with their distance scale.

For another sub or related class of events, X-ray flashes (XRFs, see Heise et al. 2001), several representatives have been detected but only one apparent spectroscopic redshift has been obtained to date, for XRF 020903 at $z = 0.25$ (Soderberg et al. GCN 1554). Others have manifest optical afterglows (Dullighan et al. 2003), with one additional localized to the vicinity of its probably host galaxy, GRB 020427 (Fruchter et al. 2002). Thus XRFs comfortably appear to be cosmological events similar to GRBs, but the relationship is nascent. Elucidation of XRFs by Swift may be problematic since the peak in their spectral energy distribution tends to fall below the Burst Alert (BAT) telescope's threshold (Fenimore 2003; Band 2003).

For short bursts (durations < 2 s) we presently have no information informing us of their luminosity and distance distributions. The optical afterglows of short bursts are predicted to be at least ten times fainter than those of long bursts (Panaitescu, Kumar, & Narayan 2001) and thus difficult to study even when detected. However, their hard X-ray afterglows appear to have been detected in the aggregate (Lazzati, Ramirez-Ruiz, & Ghisellini 2001; Connaughton 2002). Even in the absence of optical detections, the X-ray Telescope (XRT) on Swift promise should be able to detect X-ray lines which will reveal the redshifts of the sources of short GRBs (Meszaros & Rees 2003).

REFERENCES

Amati, L., et al. 2002, A&A, 390, 81
Band, D.L. 2003, ApJ, 588, 945
Band, D.L., Norris, J.P., & Bonnell, J.T. 2004, in preparation
Bennett, C.L., et al. 2003, ApJS, 143, 97
Berger, E., et al. 2003a, ApJ, 588, 99
Berger, E., et al. 2003b, astro-ph/0307228
Blakeslee, J.P., et al. 2003, ApJ, 589, 693
Bromm, V., & Loeb, A. 2002, ApJ, 575, 111
Connaughton, V. 2002, ApJ, 567, 1028
Dullighan, A., et al. 2003, GCN notice 2326
Fenimore, E.E., & Ramirez-Ruiz, E. 2000, astro-ph/0004176
Fenimore, E.E. 2003, these proceedings
Fruchter, A., et al. 2002, GCN notice 1440
Fryer, C.L., & Meszaros, P. 2003, ApJ, 588, L25
Heise, J., et al. 2001, astro-ph/0111246
Heger, A., et al. 2003, ApJ, 591, 288
Ioka, K., & Nakamura, T. 2001, ApJ, 554, L163
Lamb, D.Q., & Reichart, D.E. 2000, ApJ, 536, 1
Lazzati, D., Ramirez-Ruiz, E., & Ghisellini, G. 2001, A&A, 379, 39
Lloyd-Ronning, N.M., & Ramirez-Ruiz, E. 2002, ApJ, 576, 101
Mackey, J., Bromm, V., & Hernquist, L. 2003, ApJ, 586, 1
Meszaros, P., & Rees, M. 2003, ApJ, 591, 91
Norris, J.P., Marani, G.F., & Bonnell, J.T. 2000, ApJ, 534, 248
Norris, J.P. 2002, ApJ, 579, 386
Panaitescu, A., Kumar, P., & Narayan, R. 2001, ApJ, 561, L171
Reichart, D.E., et al. 2001, ApJ, 552, 57
Salmonson, J.D. 2001, ApJ, 546, L29
Schirber, M., & Bullock, J.S. 2003, ApJ, 584, 110
Soderberg, A., et al. 2002, GCN notice 1554
Tonry, J.L., et al. 2003, ApJ, 594, 1
Turner, M.S. 2001, astro-ph/0108103
Woosley, S.E., & MacFayden, A.I. 1999, A&AS, 138, 499